\documentclass[aps,prl,twocolumn,amsmath,amssymb,longbibliography,superscriptaddress]{revtex4-1}
\usepackage{graphicx} 
\makeatletter
\def\amsbb{\use@mathgroup \M@U \symAMSb}
\makeatother
\usepackage{bbm,overpic}
\usepackage{amsmath,amsfonts}
\makeatletter
\def\amsbb{\use@mathgroup \M@U \symAMSb}
\makeatother
\usepackage[bbgreekl]{mathbbol}
\newcommand\ve[1]{\boldsymbol{#1}}
\newcommand{\ma}[1]{\ensuremath{\amsbb{#1}}}
\newcommand{\xin}{\ve x^{(0)}}

\newcommand{\xout}{x^{(L\!+\!1)}}

\begin{document}

\title{Finite-time Lyapunov exponents of deep neural networks}
\author{L. Storm}
\affiliation{Department of Physics, University of Gothenburg, 41296 Gothenburg, Sweden}
\author{H. Linander}
\affiliation{Department of Mathematical Sciences, Chalmers Technical University and University of Gothenburg, Sweden}
\author{J. Bec}
\affiliation{MINES Paris, PSL Research University, CNRS, Cemef, Sophia-Antipolis, France F-06900}
\affiliation{Universit\'e C\^ote d'Azur, Inria, CNRS, Cemef, Sophia-Antipolis, France, F-06900}
\author{K. Gustavsson}
\affiliation{Department of Physics, University of Gothenburg, 41296 Gothenburg, Sweden}
\author{B. Mehlig}
\affiliation{Department of Physics, University of Gothenburg, 41296 Gothenburg, Sweden}

\begin{abstract}
We compute how small input perturbations affect the output of deep neural networks, exploring an analogy between deep networks and  dynamical systems, where the growth or decay of local perturbations is characterised by  finite-time Lyapunov exponents.  We show that the maximal exponent forms  geometrical structures in input space, akin to coherent structures in dynamical systems.  Ridges of large positive exponents divide input space into  different regions that  the network associates with different classes. These ridges visualise the geometry that deep networks construct in input space, shedding light on the fundamental mechanisms underlying their learning capabilities.
\end{abstract}

\maketitle

Deep neural networks can be trained to model complex functional relationships~\cite{jumper2021highly}.
 The  expressivity  of such neural networks -- their ability to unfold intricate data structures -- increases exponentially as the number of layers increases~\cite{poole2016exponential}. 
 However, deeper networks are harder to train, 
due to the multiplicative growth or decay of signals as they propagate 
through the network.
This multiplicative amplification, also known as the unstable-gradient problem~\cite{mehlig2021machine},  causes signals to either explode or vanish in magnitude if
the number of layers is too large. 
A second important problem is that we lack insight into the learning mechanisms.
Although there is some intuition for shallow networks~\cite{MinskyPapert},
there is still no general understanding of the principles that cause some architectures 
to fail, while others work better. 
 
\begin{figure}[tb]
    \centering
    \includegraphics[scale=0.3]{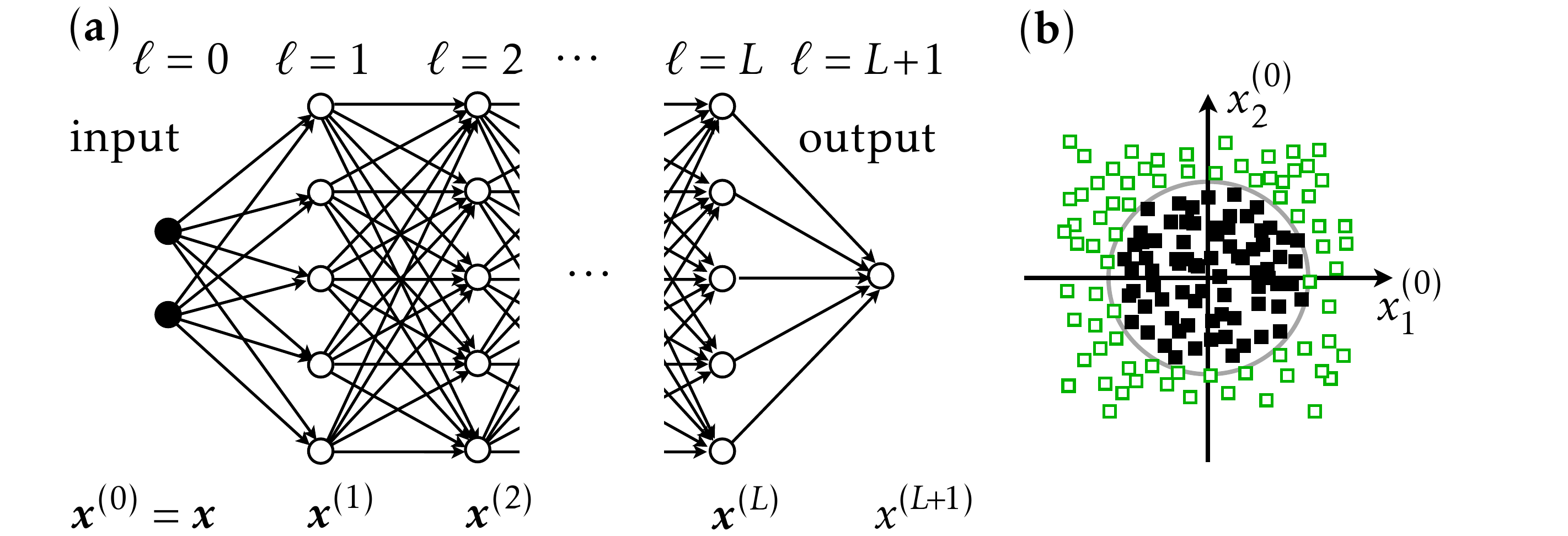}
    \caption{    \label{fig:schematic} Classification with a fully connected feed-forward network. ({\bf a}) Layout with two input components $x_1^{(0)}$ and $x_2^{(0)}$, $L$~hidden layers with $N=5$ neurons, and one output $\xout$ for classification. ({\bf b})~Two-dimensional input plane (schematic) for a classification problem with a circular
    decision boundary that separates input patterns with  targets $t=+1$ ($\blacksquare$) from those with  $t=-1$ ($\Box$, green). 
    %The data set contains $4\times10^4$ input patterns.  The network parameters are determined 
  %  by minimising the output  error $[\xout-t(\xin)]^2$ using 90\% of the data. The remaining 10\% form a test set.
    }
\end{figure}

For a common type of deep networks, the so-called multi-layer perceptrons [Fig.~\ref{fig:schematic}({\bf a})], we show that these two problems are closely related.
 We  exploit the fact that such networks are discrete dynamical systems; inputs $\xin$ are mapped iteratively through 
 $ x_i^{(\ell)} = g( \sum_{j=1}^{N_\ell} w_{ij}^{(\ell)} x_j^{(\ell-1)}\!-\theta_i^{(\ell)})$. Here, $g(\cdot)$ is a non-linear activation function \cite{mehlig2021machine}, the layer index $\ell=0,\ldots,L\!+\!1$ plays the role of time, $L$ is the number of hidden layers,   
 $N_\ell$ is the number of neurons in layer $\ell$, and the weights $w_{ij}^{(\ell)}$ and thresholds $\theta_i^{(\ell)}$ are  parameters.
Sensitivity of   $\ve x^{(\ell)}$ to small changes in
the inputs $\xin=\ve x $ corresponds to exponentially growing perturbations in a chaotic system with positive maximal Lyapunov exponent  \cite{ott2002chaos,chaosbook} $\lim_{\ell\to \infty}  \lambda_1^{(\ell)}(\ve x)$, with growth rate $\lambda_1^{(\ell)}(\ve x) = \ell^{-1}\log |\delta\ve x^{(\ell)}|/|\delta\ve x|$. 
The multiplicative ergodic theorem~\cite{ott2002chaos} guarantees 
that $\lambda_1^{(L)}(\ve x)$ converges as $L\to\infty$ to a limit that is independent of $\ve x$.
   
   The standard way of initialising network parameters is to choose zero thresholds and random weight matrices with independent Gaussian-distributed elements with
zero mean, and variance $\sigma_w^2$. In this case, the Lyapunov exponents are determined by a product of random matrices \cite{crisanti1993products}, and in the
mean-field limit  of $N_\ell =N\to\infty$, one finds $\lambda_1^{(L)} \sim \log( G N \sigma_w^2)$ independent of $\ve x$ and $L$. Here, 
the constant $G$ depends on the choice of activation function~\cite{pennington2017resurrecting}.
  This  relation explains why  the initial weight variance should be chosen so  that $G N \sigma_w^2 =1$, because then signals neither contract nor expand \cite{pennington2017resurrecting,SutskeverMartensDahlHinton,GlorotBengio2010}, stabilising the learning. 
  The maximal Lyapunov exponent also determines the success or failure in predicting chaotic time series with
recurrent networks \cite{jaeger2004harnessing,pathak2018model,storm2022constraints} that use large reservoirs of  neurons with random weights.
In that case,
the mean-field limit $N\to\infty$ works very well~\cite{storm2022constraints}. 
% More generally, the spectrum of all Lyapunov exponents correlates with training success of recurrent networks~\cite{vogt2000lyapunov}.

 For finite  $N$ and $L$, the maximal finite-time Lyapunov exponent (FTLE)  $\lambda_1^{(L)}(\ve x)$  depends on the input $\ve x$. Averaging over input patterns yields  an
estimate for the Lyapunov exponent~\cite{vogt2000lyapunov}, but even if the average $\langle \lambda_1^{(L)}(\ve x)\rangle$ over  inputs vanishes, 
some patterns may exhibit
large positive exponents,  causing the  training to fail. 

Moreover,   the weights of a trained network are not random but should reflect what the network has learned about the inputs.  This raises the question: does
the maximal exponent form  geometric structures in input space, just as in  dynamical systems where the ridges of high FTLE  define
Lagrangian coherent structures~\cite{haller2000lagrangian,lucarini2017edge,beneitez2020edge}?  How 
does the variations of the maximal FTLE in input-space depend on the number $L$ of layers of the network, and 
on its width $N$?

  To answer these questions, we computed the maximal FTLEs for
 fully connected deep neural networks with different widths and numbers of layers.
 For a simple classification problem with two-dimensional inputs $\ve x$ divided into two classes with targets~$t(\ve x)= \pm 1$ [Fig.~\ref{fig:schematic}({\bf b})], we show how the $\ve x$-dependence of the maximal FTLE changes when changing $N$ and $L$. For narrow networks (small $N$), 
we find that the maximal
 FTLE forms ridges of large exponents in the input plane, much like Lagrangian coherent structures in high-dimensional dynamical systems~\cite{haller2000lagrangian,lucarini2017edge,beneitez2020edge}.
 These ridges provide insight into the learning process, illustrating how the network learns to change its output
by order unity in response to a small shift of the input pattern across
the decision boundary. 
However, as the network width grows, we
see that the ridges disappear, 
suggesting a different learning mechanism.  Similar conclusions
 hold for a more complex classification problem using  the MNIST data set of hand-written digits, where 
 FTLE structures in input space explain variations
 in classification accuracy and predictive uncertainty.

{\em Finite-time Lyapunov exponents}. Figure~\ref{fig:schematic}({\bf a}) shows a multi-layer perceptron~\cite{mehlig2021machine}, a fully-connected feed-forward neural network with $L$ hidden layers, $N_0$ input components, $N$ neurons per hidden layer with  non-linear activation functions, and $N_{L+1}$ output neurons. The network maps every 
input $\xin=\ve x $ to an output $\xout $. Weights and thresholds are varied to minimise the output error $[\xout-t(\ve x )]^2$, so that 
the network predicts the correct target $t(\ve x)$ for each input $\ve x$.
 The sensitivity of 
$\ve x^{(\ell)}$ 
to small changes $\delta \ve x$ is determined by linearisation,
\begin{equation}
\label{eq:product}
   \delta\ve{x}^{(\ell)} \!=\!\ma D^{(\ell)} \ma W^{(\ell)}\!\cdot\!\cdot \!\cdot\,\ma D^{(2)} \ma W^{(2)}\ma D^{(1)} \ma W^{(1)}\delta{\ve x}\!\equiv\! {\ma J}_\ell\delta{\ve x}\,.
\end{equation}
Here,  $\ma W^{(\ell)}$ are the weight matrices, and $\mathbb{D}^{(\ell)}$ are diagonal matrices with elements $D_{ij}^{(\ell)}=g'(b_i^{(\ell)}) \delta_{ij}$, with
$b_i^{(\ell)} = \sum_{j=1}^{N_\ell} w_{ij}^{(\ell)} x_j^{(\ell-1)}\!-\theta_i^{(\ell)}$ and $g'(b_i^{(\ell)}) = \tfrac{\rm d}{{\rm d}b}g(b)|_{b=b_i^{(\ell)}}$.
The Jacobian ${\ma J}_\ell(\ve x)$ characterises the growth or decay 
of small perturbations to $\ve x$~\cite{ott2002chaos,chaosbook}. 
Its maximal singular value $\Lambda_1^{(\ell)}(\ve x)$ increases or decreases exponentially as a function of $\ell$, with rate $\lambda_1^{(\ell)}(\ve x) \equiv  \ell^{-1}\log \Lambda_j^{(\ell)}(\ve x)$. The singular values $\Lambda_1^{(\ell)}(\ve x) >\Lambda_2^{(\ell)}(\ve x)>\ldots$  are the square roots of the 
non-negative eigenvalues of the right Cauchy-Green tensor $\ma J_\ell^\top(\ve x)\ma J_\ell(\ve x)$.
The maximal eigenvector of  $\ma J_\ell^\top(\ve x) \ma J_\ell(\ve x) $ determines the direction of maximal stretching, i.e. in which  input direction the output changes the most, starting from a given input $\ve x$.

FTLEs and Cauchy-Green tensors
are used in solid mechanics to identify elastic deformation patterns~\cite{truesdell65nonlinear},
and to find regions of instability in  plastic deformation~\cite{ren2012plastic}  and crack initiation~\cite{jin2021failure}.
More generally, FTLEs help to characterise the sensitivity of complex dynamics to
initial conditions~\cite{vannitsem2017predictability,uthamacumaran2021review,morozov2020long}.
In fluid mechanics, they explain
the alignment of particle transported by the fluid~\cite{ni2014alignment,bezuglyy2009poincare}, providing valuable insight into the stretching and rotation of fluid elements over time and space~\cite{johnson2017analysis}. FTLEs allow to  identify Lagrangian coherent structures~\cite{haller2000lagrangian,lucarini2017edge,beneitez2020edge};  strongly repelling fluid-velocity structures that help to organise and understand flow patterns~\cite{gibson2008visualizing}. These geometrical structures appear as surfaces of large maximal FTLEs, orthogonal to the maximal
stretching direction.

 In applying these methods to neural networks, one should recognise several facts. First, in deep neural networks, the weights change from layer to layer. Therefore the corresponding dynamical system is not autonomous. Second, the number $N_\ell$ of neurons per layer may change as a function of $\ell$, corresponding to a changing phase-space dimension.  Third, the neural-network weights are trained. This limits the exponential growth of the maximal singular value, as we show below.
 Fourth, one can use different activation functions, such as the piecewise linear ReLU function \cite{mehlig2021machine}, or the smooth tanh function~\cite{pennington2017resurrecting}.
 Here we use $g(b) = \mbox{tanh}(b)$, 
 so that the network map is continuously differentiable just like the dynamical systems for which Lagrangian coherent structures were found and analysed.

{\em  Two-dimensional data set.} To illustrate the geometric structures formed by the maximal FTLE, we first consider a toy problem. 
The data set [Fig.~\ref{fig:schematic}({\bf b)}] comprises
$4\times10^4$ input patterns, with 90\% 
used for training, the rest for testing.
We trained fully connected feed-forward networks on this data set by stochastic gradient descent, minimising  the
output error $[\xout-t(\ve x )]^2$. In this way we obtained classification accuracies of at least 98\%. We considered different network layouts, changing the numbers of layers and hidden neurons per layer.
 The
weights were initialised as independent Gaussian random numbers with zero mean and variance $\sigma_w^2 \sim N^{-1}$, 
while the  thresholds were initially set 
to zero. After training,
we computed the maximal FTLE in layer $L$ and the associated stretching direction from Eq.~(\ref{eq:product}) as
described in Refs.~\cite{eckmann1985ergodic,okushima2003new}.

\begin{figure}[t]
    \centering
    \includegraphics[width=\columnwidth]{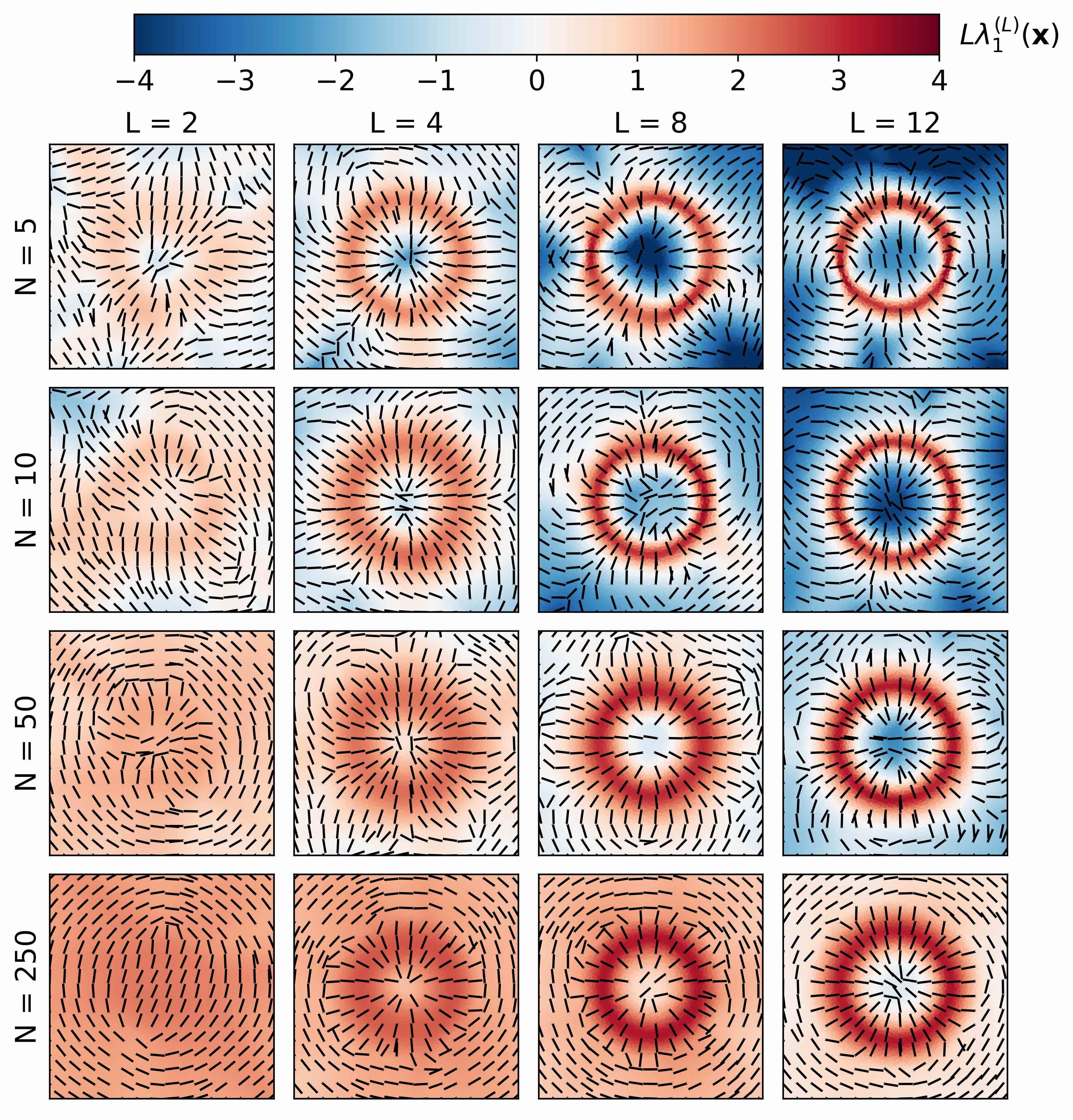}
    \caption{ \label{fig:FTLE_field} Geometrical FTLE structures in input space for different widths $N$ and depths $L$ of fully-connected feed-forward neural networks trained 
    on the data set shown schematically  in Fig.~\ref{fig:schematic}({\bf b}). Shown is the colour-coded magnitude of 
    $L \lambda_1^{(L)}(\ve x)$, and the  maximal stretching directions (black lines).}  
\end{figure}
\begin{figure}[t]
    \centering
    \includegraphics[width=0.9\columnwidth]{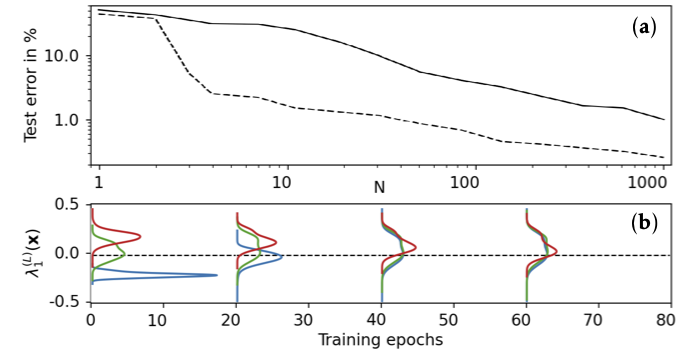}
    \caption{ \label{fig:2d} ({\bf a}) Classification error for a fully connected feed-forward network with $L=2$ hidden layers with random weights (not trained),
    and trained output weights,  as a function of the number $N$ of hidden neurons per layer  (solid black line). Also shown
    is the classification error for the fully trained network (dashed line). Both curves were obtained for the data set
    shown schematically in Fig.~\ref{fig:schematic}({\bf b}).
    ({\bf b})~Evolution of maximal-FTLE distribution as a function of training time measured in epochs~\cite{mehlig2021machine}, for a network with $L=8$  hidden layers with $N=50$ neurons per layer. The  weights were initialised with different variances,  $\log GN\sigma_w^2=-0.2$ (blue), $0$~(green), and $0.2$ (red).}
       \end{figure}
The results are summarised in Figure~\ref{fig:FTLE_field}, which shows
maximal-FTLE fields for trained networks with different layouts.  First, we see that  the ridges of large positive $\lambda_1^{(L)}({\ve x})$ align with
the decision boundary between the two classes [Fig.~\ref{fig:schematic}({\bf b})]. 
 The ridges are most prominent for small $N$
and large~$L$.  In this case, the network learns by grouping the inputs into two different 
basins of attraction for $t=\pm 1$, separated by a ridge of positive  $\lambda_1^{(L)}({\ve x})$.  
A small shift of the
input across
the decision boundary leads to a substantial
change in the output. 

Second, the contrast increases as $L$ becomes larger, quantifying the exponential expressivity of deep neural networks.  For larger $L$, the network can resolve smaller input distances $\delta\ve x$ because the singular values  increase/decrease exponentially from layer to layer.  Comparing networks of different
depths, we find that  $L \lambda_1^{(L)}(\ve x)$ saturates for large $L$,  on the ridge. This is a consequence of the
training: the network learns to produce output differences on the order of $\delta x^{(L+1)} \sim 1$, and to resolve input differences $\delta \ve x$ on the scale of
the mean distance between neighbouring patterns over the decision boundary. 
 Therefore, the saturation value is larger when the
number densities of input pattern is higher (not shown).
Even though $\Lambda_1^{(L)}(\ve x)$ saturates,  $\Lambda_2^{(L)}(\ve x)< 1$ decreases exponentially as $L$ grows (not shown), thereby causing 
$\Lambda_1^{(L)}(\ve x)/\Lambda_2^{(L)}(\ve x)$ to increase exponentially, 
as in dynamical systems~\cite{haller2011variational}.

 Third, the ridges gradually disappear as the number $N$  of hidden neurons per layer increases,
 because the maximal singular value of $\ma J_L(\ve x)$ approaches a definite $\ve x$-independent limit as $N\to\infty$ at fixed $L$.
In the infinite-width limit, training is equivalent to kernel regression with
a kernel that is  independent of the inputs in the training data set \cite{chen2020label,jacot2018neural}.  
% This is consistent with the disappearance of the ridges when $N\to\infty$.
But how can the network distinguish inputs with different targets in this case, without ridges indicating decision boundaries?  One possibility is that  the large number of  hidden neurons allows the network to embed the inputs into a high-dimensional space where they can be separated thanks to the universal approximation theorem~\cite{kidger2020universal}. In this case, 
training only the output weights (and threshold) should suffice. Figure~\ref{fig:2d}({\bf a}) confirms this, as the 
classification error decreases with increasing embedding dimension, for random hidden weights.  
%Since the network classifies by random embedding, there is no need to train the hidden neurons. 
We remark that the classification error of the fully trained network is  smaller than the error with random hidden weights. This is not surprising, since different random embeddings have different classification errors when the number of patterns exceeds twice the embedding dimension~\cite{cover1965geometrical,mehlig2021machine}.
 
Fourth, Figure~\ref{fig:FTLE_field} also shows the maximal stretching directions.  For large $L$ they become orthogonal to the ridges of large $\lambda_1^{(L)}(\ve x)$.
This demonstrates
that there is a stringent analogy between the FTLE ridges of deep neural networks and Lagrangian coherent structures. 
The stretching patterns
%  in this simple two-dimensional system 
appear to exhibit singular points 
where the maximal stretching tends to zero \cite{wilkinson2009fingerprints}, reflecting topological constraints  imposed on the direction field.

Fifth, one may wonder how the FTLE structures depend on %the
weight initialisation. When weights are initialised 
with a small variance, $\sigma_w \ll 1$,
most FTLEs are negative initially [blue in Fig.~\ref{fig:2d}({\bf b})].   This implies a slowing down of the initial training 
(vanishing-gradient problem). To see this, consider the fundamental forward-backward dichotomy of deep neural networks \cite{mehlig2021machine}: weight updates in the stochastic-gradient algorithm are given by
$\delta w_{mn}^{(\ell)}\propto \Delta_m^{(\ell)} x_n^{(\ell-1)}$, where
\begin{equation}
\label{eq:errors}
[\ve \Delta^{(\ell)}]^{\sf T}  =  [\ve \Delta^{(L)}]^{\sf T} \ma D^{(L)} \ma W^{(L)}\cdots  \ma D^{(\ell\!+\!1)} \ma W^{(\ell\!+\!1)} \ma D^{(\ell)} \,,
\end{equation}
and
$\Delta^{(L)}_j =  g'(b^{(L+1)} ) [x^{(L+1)}\!- \!t(\ve x)]  g'(b_j^{(L)} ) w_j^{(L+1)} $.
It follows from Eq.~(\ref{eq:errors}) that  negative FTLEs cause
small weight increments $\delta w_{mn}^{(\ell)}$. Conversely, 
when the maximal FTLE is positive and too large, the weights grow rapidly, leading to training instabilities. 
Remarkably, Figure~\ref{fig:2d}({\bf b}) demonstrates a self-organising effect due to
training: the distributions of the maximal FTLE converge to centre around zero. 
This is explained by the fact that the network learns by creating maximal-FTLE ridges in input space: to accommodate positive and negative $\lambda_1^{(L)}(\ve x)$, the
distribution centres around zero, 
alleviating
the unstable-gradient problem. 

  \begin{figure}[tb]
    \includegraphics[width=0.9\columnwidth]{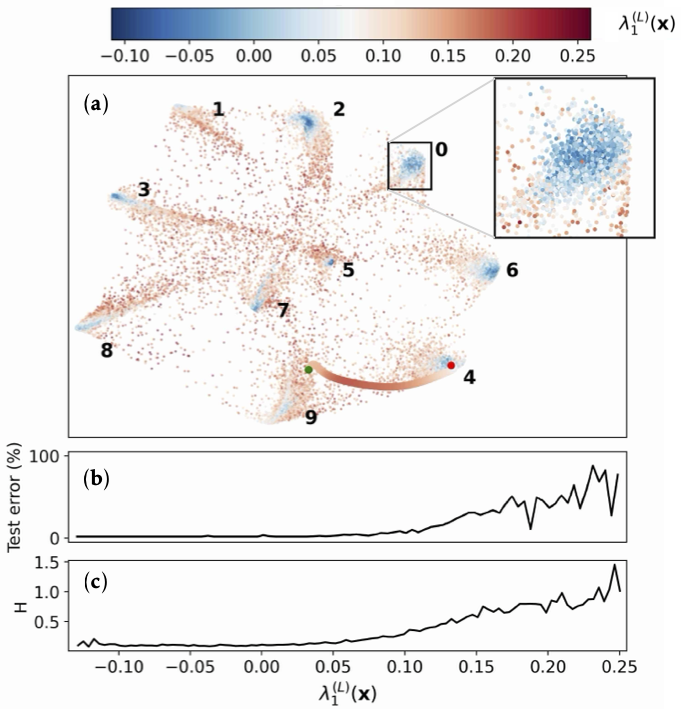}
    \caption{  \label{fig:MNIST}
Maximal-FTLE field for the MNIST data \cite{MNIST}. A fully connected feed-forward network with  $N=20$ neurons per hidden layer, $L=16$  hidden
layers, and a softmax  layer with ten outputs was trained to a classification accuracy of 98.88\%. The maximal FTLE was calculated for each of the $28^2$-dimensional inputs and projected 
to two dimensions (see text).  ({\bf a})  Training data in the non-linear projection. 
  For each input, the maximal FTLE $\lambda_1^{(L)}$ is shown colour-coded (legend).  The box contains 93\%  of the recognised digits $0$.
  A  threefold blow up of this box is also shown. The  line represents a sequence of adversarial attacks from $9$ to $4$ (see text), with $\lambda_1^{(L)}(\ve x)$ colour-coded.
({\bf b}) Classification error on the test set as a function of $\lambda_1^{(L)}(\ve x)$. 
({\bf c})~Predictive uncertainty $H$ (see text) as a  function of $\lambda_1^{(L)}(\ve x)$.
 }
  \end{figure}

{\em  MNIST data set.} This data set consists of 60,000 images of handwritten digits $0$ to $9$. Each grayscale image has $28\times 28$ pixels and was pre-processed to facilitate machine learning \cite{MNIST}. Deep neural networks can achieve  high precision in classifying this data,
with accuracies of up to $99.77$\% 
on a test set of 10,000 digits \cite{ciregan2012multi}. 

We determined the maximal-FTLE field for this data set  for a 
network with $L=16$ hidden layers, each containing
$N=20$ neurons, 
and a standard softmax layer with ten outputs \cite{mehlig2021machine}. %The network was trained to a classification accuracy of 98.88\% on the training set. 
To visualise the geometrical structures in the $28^2$-dimensional input space, we projected it to two dimensions as follows. We added a bottleneck layer with two neurons to the fully trained network, just before the softmax-output layer. We retrained only the weights and  thresholds of this additional layer and the output layer, keeping all other hidden neurons unchanged. The local fields 
$b_1$ and $b_2$ of the two bottleneck neurons
are the coordinates of the two-dimensional representation shown in Figure~\ref{fig:MNIST}({\bf a}). We see that the input data separate 
into ten distinct clusters corresponding to the ten digits. 
The maximal  FTLEs at the centre of these clusters are very small or even negative,  indicating that the output is not sensitive to small input changes. 
These regions are delineated by areas with significantly
larger  positive FTLEs [see $3\times$zoom in panel~({\bf a})].
Figure~\ref{fig:FTLE_field} leads us to expect that patterns with large $\lambda_1^{(L)}(\ve x)$ {are located near the decision boundaries in high-dimensional input space. 
This is verified by strong correlations between $\lambda_1^{(L)}(\ve x)$ and both the classification error and the predictive uncertainty.
Figure~\ref{fig:MNIST}({\bf b}) shows  that the classification error on the test
 set  is larger for inputs $\ve x$ with larger  $\lambda_1^{(L)}(\ve x)$. Figure~\ref{fig:MNIST}({\bf c})  shows that large values of $\lambda_1^{(L)}(\ve x)$ correlate with high predictive uncertainty, measured by the entropy $H$ of the posterior predictive distribution~\cite{gawlikowski2021}. For softmax outputs, where $x_i^{(L+1)}$ %that
 can be interpreted as probabilities,   $H = -\sum_i \langle x_i^{(L+1)}\rangle \log \langle  x_i^{(L+1)}\rangle$, where $\langle \cdot\rangle$ denotes an average over an ensemble of networks  (here with ten members)  with the same layout but different weight initialisations~\cite{lakshminarayanan2017,hoffmann2021}. These observations confirm that ridges of maximal FTLEs localise the decision boundaries.
 
Figure~\ref{fig:MNIST}({\bf a}) also shows $\lambda_1^{(L)}(\ve x)$ along a path generated by an adversarial attack. The attack begins
from a sample within
the cluster corresponding to the digit $9$ and aims to transform it into a digit $4$ by making small perturbations to the input data~\cite{madry2018} toward class $4$.
We see that the maximal FTLE is small at first, then increases as the path approaches
the decision boundary, and eventually decreases again.
 This indicates that our conclusions regarding the correlations between large maximal FTLEs and decision boundaries extend to neighbourhoods of the MNIST training set that contain
 patterns the network has not encountered during
 training.

 {\em Conclusions.} 
We explored geometrical structures formed by the maximal FTLE in input space,  for deep
neural networks trained on different classification problems. We found that ridges of positive exponents
define  the decision boundaries, for a two-dimensional toy classification problem, and for a high-dimensional data set of hand-written digits.}
In the latter case, we projected the high-dimensional
input space to two dimensions, and found
that the network maps digits into distinct clusters surrounded by FTLE ridges at the decision boundaries.
This conclusion is supported by the fact that the locations of the FTLE ridges correlate with low classification accuracy and high predictive uncertainty.
%Our analysis also showed that 

The network layout determines how prominent the FTLE structures are.  
As the number of layers increases, the ridges sharpen, emphasising their role in learning and classification.
However, as the number of hidden neurons per layer tends to infinity, the FTLE structures  disappear.
 In this limit, the network separates the inputs  by embedding them into a high-dimensional space, rendering training of the hidden neurons unnecessary. 

It is important to underscore that the two different ways to learn, by FTLE ridges or 
embedding, result in qualitative differences regarding classification errors and predictive uncertainties, and may also affect how susceptible a network is to adversarial attacks. The geometrical method presented here extends to other network architectures (such as convolutional networks), and will help to visualise and understand the mechanisms that allow such neural networks to learn.

\begin{acknowledgments}
LS was supported by grants from  the Knut and Alice Wallenberg (KAW) Foundation (no. 2019.0079) and Vetenskapsr\aa{}det  (VR), no.~2021-4452.
JB received support from UCA-JEDI Future Investments (grant no. ANR-15-IDEX-01).
HL was supported by  a grant from the KAW Foundation. KG was supported by VR grant~2018-03974, and BM by  VR grant~2021-4452.
Part of the the numerical computations for this project were performed on resources provided by the Swedish National Infrastructure for Computing (SNIC).
\end{acknowledgments}

%\bibliography{references}

\end{document}